\documentclass[12pt,letterpaper]{JHEP3}
\usepackage{amsmath}
\usepackage{cite}
\usepackage[vcentermath]{youngtab}
\usepackage{epic}

\newdimen\tableauside\tableauside=1.0ex
\newdimen\tableaurule\tableaurule=0.4pt
\newdimen\tableaustep
\def\phantomhrule#1{\hbox{\vbox to0pt{\hrule height\tableaurule width#1\vss}}}
\def\phantomvrule#1{\vbox{\hbox to0pt{\vrule width\tableaurule height#1\hss}}}
\def\sqr{\vbox{%
  \phantomhrule\tableaustep
  \hbox{\phantomvrule\tableaustep\kern\tableaustep\phantomvrule\tableaustep}%
  \hbox{\vbox{\phantomhrule\tableauside}\kern-\tableaurule}}}
\def\squares#1{\hbox{\count0=#1\noindent\loop\sqr
  \advance\count0 by-1 \ifnum\count0>0\repeat}}
\def\tableau#1{\vcenter{\offinterlineskip
  \tableaustep=\tableauside\advance\tableaustep by-\tableaurule
  \kern\normallineskip\hbox
    {\kern\normallineskip\vbox
      {\gettableau#1 0 }%
     \kern\normallineskip\kern\tableaurule}%
  \kern\normallineskip\kern\tableaurule}}
\def\gettableau#1 {\ifnum#1=0\let\next=\null\else
  \squares{#1}\let\next=\gettableau\fi\next}

\title{Holography and entropy bounds\\ in the plane wave matrix model}

\author{%
Raphael Bousso and Aleksey L. Mints \\
Center for Theoretical Physics, Department of Physics\\
University of California, Berkeley, CA 94720-7300, U.S.A.\\
{\em and}\\
Lawrence Berkeley National Laboratory, Berkeley, CA 94720-8162, U.S.A.\\
E-mail: \email{bousso@lbl.gov, mints@socrates.berkeley.edu}}

\abstract{%
  As a quantum theory of gravity, Matrix theory should provide a
  realization of the holographic principle, in the sense that a
  holographic theory should contain one binary degree of freedom per
  Planck area.  We present evidence that Bekenstein's entropy
  bound, which is related to area differences, is manifest in the
  plane wave matrix model.  If holography is implemented in this way,
  we predict crossover behavior at strong coupling when the energy
  exceeds $N^2$ in units of the mass scale.}

\preprint{\hepth{0512201} \\ UCB-PTH-05/32 \\ LBNL-58947}

\begin{document}

\section{Introduction}

The holographic principle~\cite{Tho93,Sus95,CEB2} requires that the
surprising entropy bounds apparent in nature~\cite{CEB1} should be
manifest in quantum gravity.  For asymptotically Anti-de~Sitter
spacetimes this was realized by the Maldacena
conjecture~\cite{Mal97} in which $A$ binary degrees of freedom of an
$SU(N)$ Yang-Mills theory suffice to describe a region of surface area
$A$~\cite{SusWit98}, in Planck units.  This holographic behavior
follows from a UV/IR relation that places a UV cutoff on the field
theory as the size of bulk regions is decreased~\cite{PeePol98}.

When the bulk region becomes as small as the AdS curvature scale, the
field theory reaches its lowest possible UV cutoff.  Only the
zero-modes on the $S^3$ are retained, and the theory reduces to matrix
quantum mechanics.  At this point holography is still manifest
with $N^2$ degrees of freedom describing a region of area $N^2$.  But
for smaller regions it is not clear how to eliminate any further
degrees of freedom from the field theory.  For similar reasons
holographic state counting is quite obscure in the matrix models
proposed to describe M-theory on asymptotically flat space~\cite{BFSS}
or on the plane wave in eleven dimensions~\cite{BMN}.

In this paper, we propose that the entropy bound that becomes manifest
in matrix quantum mechanics is not the covariant bound relating
entropy to surface area.  Rather, it is the Bekenstein
bound~\cite{Bek81}, which states that the entropy of a system will not
exceed its mass times its linear size.  We find evidence that this
bound is saturated by the matrix degrees of freedom available for the
description of weakly curved geometries.

The Bekenstein bound appears to be connected to the holographic
principle in that it arises from the generalized covariant entropy
bound~\cite{FMW} in a certain weak-gravity limit~\cite{Bou03}.  In its
regime of validity it is much tighter than the covariant bound.  Thus,
it may govern the emergence of local weakly gravitating regions from a
more fundamental theory~\cite{Bou04}, similar to the more general role
envisaged for the covariant bound.  However, it has suffered from
ambiguities in its definition of entropy (recent discussions of this
problem include
Refs.~\cite{Pag00c,Bek00b,Bou03a,MarRoi04,MarMin03,Pag04,Bek04}).

Attempts to rectify this situation led to the proposal~\cite{Bou03b}
to formulate Bekenstein's bound in terms of the discrete light-cone
quantization (DLCQ) of theories on backgrounds with a null Killing
vector: The entropy in a sector with $N$ units of momentum should not
exceed $2\pi^2N$:\footnote{In Ref.~\cite{Bou03} the Bekenstein bound
  is derived as the integrated loss of cross-sectional area of
  parallel light-rays focused by a matter system (see also
  Ref.~\cite{BouFla03,StrTho03}).  It arises not in its original form
  ($S\leq 2\pi M R$, where $R$ is the largest linear dimension), but
  in the stronger form $S\leq \pi M \Delta x$, where $\Delta x$ is the
  width of the system along an arbitrary direction in its rest frame.
  It is more natural to work in the lightcone frame picked out by the
  light-sheet.  One then has $S\leq\pi P_- \Delta x^-$, where the null
  coordinate $x^-$ and the longitudinal momentum $P_-$ are defined as
  usual~\cite{Bou03b}.  Ambiguities in defining the spatial extent of
  quantum states can be suppressed by compactifying the null
  direction.  Then the momentum is quantized ($P_- = 2\pi N/\Delta
  x^-$), and the bound becomes $S\leq 2\pi^2 N$.  Here $S$ is the
  logarithm of the number of discrete states~\cite{Bou03a} in the
  sector with $N$ units of momentum.}
\begin{equation}
S\leq 2\pi^2 N~.
\label{eq-bek}
\end{equation}

It is natural to test this proposal in the context of M-theory.
M-theory is {\em defined\/} in terms of its DLCQ as a $U(N)$
matrix model~\cite{Sus97}.  Moreover, we expect that it represents a
consistent quantum theory of gravity.  This means that violations, if
found, cannot be ascribed to an artificial choice of Lagrangian.

In particular, we shall study M-theory on the eleven-dimensional plane
wave background.  Unlike Matrix theory in flat space~\cite{BFSS}, the
spectrum of the plane wave matrix model is discrete.  It is known
exactly~\cite{DSV1} at large boosts $\mu$, where the quantum mechanics
is weakly coupled.  However, the curvature of the gravity dual is
strong in this regime, and the Bekenstein bound is not expected to
apply.

At small $\mu$, when the plane wave is weakly curved, the matrix model
is strongly coupled.  However, there are still infinite towers of
protected states.  Without a cutoff on the lightcone energy the
protected states alone would seem to contribute an infinite amount of
entropy at any value of $N$, apparently violating the Bekenstein
bound.

However, both the full spectrum at weak coupling and the spectrum of
protected states at arbitrary coupling undergo crossover behavior
when the energy in units of $\mu$ becomes of order $N^2$, i.e., for
\begin{equation}
  E_{\rm cross}\sim N^2~.  
\end{equation}
The entropy of the full spectrum (at weak coupling) behaves as
\begin{equation}
  S\sim E~~(E\lesssim N^2)~;~~~ \qquad \qquad
  S\sim N^2\log\frac{E}{N^2}~~(E\gtrsim N^2)~.
\label{eq-prefull}
\end{equation}
The entropy of the protected states behaves as
\begin{equation}
  S_{\rm p}\sim (EN)^{1/3}~~(E\lesssim N^2)~;~~~~~
  S_{\rm p}\sim N\log\frac{E}{N^2}~~(E\gtrsim N^2)~.
\label{eq-preprot}
\end{equation}

At strong coupling, where the Bekenstein bound should apply, only the
protected states are guaranteed to be present.  We shall assume that
they dominate the entropy in the microcanonical ensemble, at least in
the vicinity of the crossover energy.  Then the above results show
that the system undergoes a kind of phase transition (smoothed out by
finite $N$) at $E\sim N^2$, both at strong and at weak coupling.  On
the gravity side, one expects this type of behavior to be associated
with a nonperturbative modification of the background~\cite{AhaMar03}
analogous to the Hawking-Page transition in Anti-de~Sitter
space~\cite{HawPag83,Wit98a}.

It would be interesting to develop a concrete proposal for the gravity
interpretation of this transition in the 11D plane wave.  No black
hole solutions asymptotic to the 11D plane wave have been found, and
it is possible that the eleven-dimensional spacetime interpretation
breaks down entirely at the crossover energy.

States associated with a strong perturbation of the background do not
contribute to the Bekenstein bound.  Thus, under the stated
assumptions, the entropy entering the bound is at most the crossover
entropy,
\begin{equation}
S_{\rm p, cross} \sim N~. 
\end{equation}
Comparison with Eq.~(\ref{eq-bek}) finds the Bekenstein bound
saturated, but not exceeded.

This paper is organized as follows.  In Sec.~\ref{sec-weak} we
consider the plane wave matrix model at weak coupling.  Adapting
general arguments for a Hagedorn/deconfinement phase transition in
weakly coupled Yang-Mills
theory~\cite{Sun00,AhaMar03,AhaMar04,AhaMar05a,AhaMar05b,Sus97,
LiMar98,HadRam04,Sem04,FurSch03}, we find crossover behavior at $E\sim N^2$,
Eq.~(\ref{eq-prefull}).  At this point the energy becomes large enough
to excite all matrix degrees of freedom; the finite,
quantum-mechanical nature of the finite $N$ system is revealed, and is
reflected in a much slower growth of entropy.

Because this result applies in a regime of strong curvature, it is not
directly relevant to the Bekenstein bound.  However, it lends
additional support to the evidence we find in Sec.~\ref{sec-encp} for
a similar transition at strong coupling.  We consider the spectrum of
protected multiplets.  It receives no corrections at strong coupling
when the model is conjectured to describe a weakly curved background.
We find that it also undergoes a transition at $E\sim N^2$, as shown
in Eq.~(\ref{eq-preprot}).

Our results are suggestive but not conclusive.  In
Sec.~\ref{sec-discuss} we clarify the additional assumptions required
to interpret the crossover entropy $S_{\rm p, cross}\sim N$ of
protected states as a manifestation of the Bekenstein bound.  In
particular, one must assume that the crossover seen for protected
states is representative of the full theory at strong coupling.

In Appendix~\ref{sec-n1} we discuss a subtlety arising for the center
of mass degrees of freedom.  Appendix~\ref{sec-protected} summarizes
properties of protected states.

\section{Entropy and crossover in the free theory}
\label{sec-weak}

In this section we introduce the plane wave matrix model and discuss
the crossover behavior of the entropy at energy $N^2$ at weak
coupling when the full spectrum is known exactly.

\subsection{The plane wave matrix model}
\label{sec-pwmm}

The $U(N)$ plane wave matrix model is given by the Hamiltonian
\begin{eqnarray} 
  H &=& R\ {\rm Tr}  \left( {1 \over 2} \Pi_A^2 - {1 \over 4} [X_A, X_B]^2 
    - {1 \over 2} \Psi^\top \gamma^A [X_A, \Psi] \right) \cr
  &+& {R \over 2} {\rm Tr}  \left( \left({\mu\over 
        3R}\right)^2  X_i^2 +  \left({\mu \over 6R}\right)^2 X_a^2 
  \right. \cr
  && \qquad \qquad \left. + 
  i {\mu \over 4R} \Psi^\top \gamma^{123} \Psi  + i {2\mu \over 3R} 
  \epsilon^{ijk} X_i X_j X_k \right)~.   
\label{eq-model}
\end{eqnarray} 
Indices $A\ldots$ run from 1 to 9; $i\ldots$ run from 1 to 3; and
$a\ldots$ run from 4 to 9.  This model was proposed~\cite{BMN} to
describe M-theory on the maximally supersymmetric plane wave
background of eleven-dimensional supergravity,
\begin{eqnarray} 
  ds^2 &=& - 2 dx^+ dx^- + (dx^A)^2 - \left( {\mu^2 
      \over 9} x^i x_i +  {\mu^2 \over 36} x^a x_a\right) (dx^+)^2~, \\
  F_{123+} &=& \mu~.
\label{eq-metric}
\end{eqnarray}

The parameter $\mu$ in the metric is a coordinate artifact; it can be
set to any value by rescaling $x^\pm$.  The M-theory limit of the
matrix model (\ref{eq-model}) is obtained by taking $N\to\infty$ while
holding $N/R$ fixed.  In this limit the model must also become
independent of $\mu$.

At finite $N$ the matrix model is expected to describe the DLCQ of
M-theory, in the sector with $N$ units of longitudinal
momentum~\cite{Sus97}.  In this case, the coordinate $x^-$ is
periodically identified with period $2\pi R$.  For finite $N$ the
boost-invariant quantity $\mu/R$ is a physical parameter that
distinguishes qualitatively different regimes.

To see this, it is useful to think of $\mu$ as the curvature radius of
the transverse dimensions as measured in the frame in which the
periodically identified hypersurfaces have spatial distance $R$.  To
obtain a good geometric description we require both $\mu\ll 1$ and
$R\gg 1$; hence, $\mu/R\ll 1$.  For large $\mu/R$ the matrix model
does not correspond to a classical background in any frame; in
particular, we do not expect the Bekenstein bound to apply.

In the limit $\mu\to\infty$ at fixed $N$ and $R$ the plane wave
matrix model (\ref{eq-model}) becomes free.  For each partition of $N$
there is a superselection sector with its own $1/2$ BPS vacuum,
corresponding to a collection of concentric fuzzy spheres~\cite{BMN}.
Let us focus on the $X=0$ sector given by the trivial partition
$N=1+1+\ldots+1$.  As we shall see, it exhibits the most rapid growth
of entropy.

\subsection{$X=0$ sector}

The $X=0$ sector has Hamiltonian
\begin{equation}
H = \frac{\mu}{3} \sum_i A^{\dagger}_i A_i + \frac{\mu}{6}
\sum_a A^{\dagger}_a A_a + \frac{\mu}{4} \sum_{I\alpha}
\psi_{I\alpha}^{\dagger} \psi_{I\alpha}
\label{eq-ham}
\end{equation}
and contains rank $N$ matrix creation operators
\begin{equation}
A_i^\dagger=\sqrt{\frac{\mu}{6R}} X^i -
i\sqrt{\frac{3R}{2\mu}}\Pi^i~,~~~
A_a^\dagger= \sqrt{\frac{\mu}{12R}} X^a -
i\sqrt{\frac{3R}{\mu}}\Pi^a~,
\label{eq-ops}
\end{equation}
as well as fermionic operators $\psi_{I\alpha}$~\cite{DSV1}.

Each creation operator contributes of order $\mu$ to the lightcone
energy, so the dimensionless energy $E\equiv -P_+/\mu$ measures the
number of quanta.  Physical states must be gauge-invariant.  Hence,
they correspond to products of traces of products of creation
operators.

It is important that there is more than one kind of matrix operator.
There is one operator for each transverse direction that the system
can move in, so their number, $q$, is of order ten.  In general
matrices do not commute, so a state with $E$ quanta can be made in
$q^E$ different ways\footnote{This is a slight simplification, even
  leaving aside the issue of trace relations discussed below.  At
  least one trace must be taken, and of course multiple traces are
  allowed as well.  Thus, we have undercounted states since the
  sequence of $E$ operators can be sprinkled with traces.  However,
  this cannot multiply the number of states by more than the number of
  partitions of $E$, $p(E)\approx \exp(\sqrt{\frac{2}{3}}\pi
  \sqrt{E})$.  Hence it will correct the entropy (\ref{eq-striv}) at
  most by a subleading term of order $\sqrt{E}$.  We have also
  overcounted states because states related by cyclic exchange of the
  creation operators within a trace are not independent.  This will
  reduce the entropy at most by a term of order $\log E$, which again
  is negligible compared to Eq.~(\ref{eq-striv}).} leading to an
entropy $S\sim E\log q$.  Since $q$ is not very large, one
has~\cite{Sun00,AhaMar03}
\begin{equation}
S\sim E~.  
\label{eq-striv}
\end{equation}

This Hagedorn behavior cannot persist indefinitely.  At sufficiently
high energy the entropy is dictated by the thermodynamics of a
quantum mechanical system with about $N^2$ degrees of freedom:
\begin{equation}
S\sim N^2\log \frac{E}{N^2}~.
\label{eq-shigh}
\end{equation}
At a crossover energy of order $N^2$ the two expressions for the
entropy match.

The argument for Hagedorn growth, Eq.~(\ref{eq-striv}), breaks down
because trace relations can lead to identifications between
states~\cite{AhaMar03}.  Apparently\footnote{There is tentative direct
  evidence for this.  One can show that all single-trace states of
  length up to $N$ are independent, even for $q>1$~\cite{BouMin05a}.
  For $q=1$ all traces longer than $N$ decompose into products of
  traces.  Hence, for $q>1$ trace relations also set in at lengths
  exceeding $N$, at least for some traces.  In a typical partition of
  $E$ most traces have length of order $\sqrt{E}$, which will exceed
  $N$ once $E$ becomes larger than $N^2$.  Hence, this is the point at
  which we expect a typical state in the Hagedorn spectrum to become
  identified with other states by trace relations.  This argument
  assumes that trace relations are not dominated by relations
  involving products of more than one trace on both sides of the
  equality.} this effect becomes important only at $E\sim N^2$.

\subsection{Full spectrum}

Other sectors exhibit a less rapid growth of states.  Consider the
``irreducible'' vacuum, which corresponds to the partition of $N$ into
just one term.  It describes a single fuzzy membrane of momentum
$N/R$.  At small energies the entropy scales as $S_{\rm M2}\sim
E^{2/3}$, as it should for a 2+1 dimensional object.  Thus, it grows
more slowly than (\ref{eq-striv}) below.  Matching to the asymptotic
density of states, Eq.~(\ref{eq-shigh}), reveals that the crossover
occurs at energies of order $N^3$.

The differing crossover energies are easily understood as follows.
Crossover happens when we notice that the matrix is finite, i.e., for
energies large enough to excite all matrix degrees of freedom.  In the
trivial vacuum there are approximately $qN^2$ matrix elements.  Each
corresponds to a creation operator that increases $E$ by about 1.
Thus, with energies of order $N^2$ they can all become excited.  In
the irreducible vacuum there are again $N^2$ oscillators, but with
different energies: There are, roughly, $i$ operators with mass
$i\mu$, $1\leq i\leq N$~\cite{DSV1}.  Hence, the crossover energy is
of order $N^3$, the energy required to excite all oscillators.

As soon as we turn on any non-zero coupling ($\mu$ large but finite),
the different partitions of $N$ cease to define superselection
sectors.  All sectors mix in the microcanonical ensemble.  The entropy
will be dominated by the sector with the most rapid growth, the $X=0$
vacuum.  Hence, the transition to the thermodynamic behavior
(\ref{eq-shigh}) will set in at energies of order $N^2$.

The crossover entropy will also be of order $N^2$.  It is worth
stressing again that the weakly coupled matrix model does not admit an
interpretation as a weakly curved geometry.  Hence, we are not in a
regime where the Bekenstein bound can be tested; we have merely noted
that the model undergoes crossover behavior at energy $N^2$.

\section{Entropy and crossover of protected states}
\label{sec-encp}

In this section we discuss the spectrum of protected states, which
will be exact at all values of the coupling.  We show that it also
undergoes a transition at energy $N^2$ when the entropy is about $N$.

\subsection{Strong coupling and protected states}

In any given sector the coupling becomes strong for
\begin{equation}
\left(\frac{R}{\mu N_{\rm max}}\right)^3 N_{\rm max} > 1~,
\end{equation}
where $N_{\rm max}$ is the largest term in the partition of
$N$~\cite{DSV1}.  As $\mu$ is decreased this will happen first for
the $X=0$ vacuum, at $\mu/R\sim 1$.

The information we have about the spectrum at strong coupling comes
from quantities which are protected for arbitrary positive values of
$\mu$~\cite{DSV2}.  In particular, multiplets in the weight $(0,n,0)$
representation of the $SU(4|2)$ symmetry group are exactly protected.
The lightcone energy of states in these multiplets is of order $n\mu$:
\begin{equation}
E\sim n~;
\end{equation}
it does not receive corrections.  

Denoted by Young supertableaux, the protected representations are
rectangular with $n$ columns and 2 rows, e.g.:
\begin{equation}
{\tiny \yngSLASH(11,11)}~.
\end{equation}
The degeneracy of each multiplet is of order $n^4$.  The entire
multiplet can be obtained by acting with supersymmetry generators on
its primary states, which are associated with the $SO(6)$ transverse
directions $X^a$ (see Appendix~\ref{sec-protected}).

\subsection{$X=0$ sector}

We will first discuss the entropy of protected states in individual
superselection sectors of the free theory ($\mu\to\infty$).  As
explained in Appendix~\ref{sec-n1}, we suppress the center of mass
degrees of freedom.  Then the single-membrane sector contains neither
operators nor states of weight $(0,n,0)$.

Let us focus instead on the $X=0$ sector, given in Eqs.~(\ref{eq-ham})
and (\ref{eq-ops}).  The primaries for $(0,n,0)$ representations are
of the form
\begin{equation}
S^{a_1...a_n}~ {\rm T\!R}_N \big(A^\dagger_{a_1} \cdots A^\dagger_{a_n}\big)
|0\rangle ~.
\end{equation}
where $S^{a_1...a_n}$ is a totally symmetric traceless tensor of
$SO(6)$, and we use the notation ${\rm T\!R}_N$ to indicate that an
arbitrary number of traces may be sprinkled among the $n$ operators,
subject to the constraint that each trace contain at most $N$
operators.  (Because of the complete symmetrization, traces longer
than $N$ are guaranteed not to be independent.  In order to exclude
the $U(1)$ one should also require that each trace include at least
two operators, but this will give a negligible correction to the
entropy.)  

Ordering the traces by their length, we may represent each primary by
a Young diagram with $n$ boxes and at most $N$ rows, where each column
represents one trace.  These auxiliary diagrams do not indicate a
group representation; they simply keep track of all the different ways
one can {\em produce\/} the $(0,n,0)$ representation, and thus of its
degeneracy in the spectrum.

Hence, the number of protected representations of energy $E$ is given
by the number of restricted partitions of $E$,
\begin{equation}
  p_N(E) \approx \frac{1}{4\sqrt{3}E} 
  \exp\left[ \pi\sqrt{\frac{2E}{3}} -
    \frac{\sqrt{6E}}{\pi}\exp\left(\frac{-\pi N}{\sqrt{6E}}\right)
  \right]~.
\label{eq-pn}
\end{equation}
This formula is valid for $1\ll E\ll N^2$.  The entropy of protected
states will thus grow as
\begin{equation}
S_{{\rm p}~(X=0)} \sim 2\pi \sqrt{\frac{E}{6}} 
\end{equation}
until a crossover threshold around $E\sim N^2$, when
\begin{equation}
S_{{\rm p, cross}~(X=0)} \sim N~.
\label{eq-spc}
\end{equation}
The contribution $4\log E$ from the degeneracy of states within each
protected multiplet is negligible.

One can understand this result from a number of perspectives.
Naively, one might expect crossover behavior for $p_N(E)$ around
$E=N$ when the restriction to at most $N$ rows first becomes
nontrivial.  However, what matters for the entropy is how the
restriction acts on {\em typical\/} partitions.  For large $E$ the
vast majority of unrestricted partitions contain about $\sqrt{E}$ rows
ranging in length up to $\sqrt{E}$; that is, the Young diagram
associated to a typical partition looks very roughly like a triangle
of height and width $\sqrt{E}$.\footnote{More precisely, by computing
  expected occupation numbers of $N\to\infty$ oscillators in
  Eq.~(\ref{eq-zpn}) below one obtains the limiting curve
  $e^{x/T}+e^{y/T}=1$, where $x$ and $y$ are coordinates along the
  edges of the Young diagram, and $T=\sqrt{6E}/\pi$.  This means that
  the expected height and width of the diagram is $T\log T$.  Hence it
  will begin to exceed $N$ slightly earlier than for $T\sim N$.  This
  slows down the growth of the degeneracy but only enough to modify
  order one prefactors in the entropy.  One still has $S\sim\sqrt{E}$
  until $T\sim N$~\cite{BalDeb05}.} This means that the restriction to
at most $N$ rows will not become important until $E\sim N^2$, when a
typical partition would prefer to have more than $N$ rows.  Therefore,
we expect crossover behavior in the number of states for $E\sim N^2$,
not $E\sim N$.

Let us gain more insight into the behavior of the entropy beyond the
crossover.  The partition function for protected representations is
identical to that for $N$ bosons in a harmonic oscillator or for a
system of $N$ harmonic oscillators of frequency $1,2,\ldots, N$, with
zero point energies removed:
\begin{equation}
Z(T) = \prod_{j=1}^N \frac{1}{1-e^{-j/T}}~.
\label{eq-zpn}
\end{equation}
Computing the entropy and energy from $Z(T)$ for $1\ll T\ll N$ yields
Eq.~(\ref{eq-pn})~\cite{TraMur03}.

For $T\to\infty$ this partition function yields $S_{{\rm
    p}\,(X=0)}=2N+N\log (T/N)$ and $E\sim NT$, consistent with the
thermodynamics of a quantum mechanical system with $N$ degrees of
freedom.  Hence
\begin{equation}
S_{{\rm p}\,(X=0)}= 2N + N\log \frac{E}{N^2}~.
\label{eq-hent}
\end{equation}
The onset of this asymptotic behavior is for $T=N$, when the
temperature is large enough to excite the frequency $N$ oscillator.
This corresponds to a lower bound $E\sim N^2$ for the asymptotic
regime, consistent with the upper limit obtained for the $T<N$ regime.
The crossover entropy, $S_{\rm p, cross}\sim N$, is also consistent
with Eq.~(\ref{eq-spc}).

\subsection{Full protected spectrum}

At any finite coupling, and in particular at strong coupling, all
sectors of the free theory must be included in the microcanonical
ensemble.  So far we have considered only the protected states of the
$X=0$ vacuum ($N=1+\ldots+1$).  The other vacua, corresponding to
nontrivial partitions of $N$, will contribute additional protected
states.  

Consider a general partition of $N$, $(N_1^{M_1}\cdots N_l^{M_l})$.
In this notation $M_i$ denotes how many times $N_i$ appears in the
sum, i.e., $N=\sum_{i=1}^l M_i N_i$.  Each term $N_i^{M_i}$
corresponds to a stack of $M_i$ coincident fuzzy spheres of individual
momenta $N_i/R$.  Each stack contributes a rank $M_i$ protected matrix
creation operator.  Thus, for the purposes of protected states, the
momentum of the stack is irrelevant; only its ``height'' $M_i$ 
matters.

The counting problem is not only how $E$ can be partitioned into
traces, but at the same time how $N$ can be partitioned into stacks,
whose height controls the maximum length of traces and whose number
controls the number of different types of traces that can be
constructed.

To compute the total number of protected states it is convenient to
introduce a chemical potential $\nu$ conjugate to $N$.  Then the
problem is an extension of the canonical partition function generating
the partitions of $E$.  A stack $N_i^{M_i}$ forms in response to the
chemical potential $\nu$, in analogy to an oscillator of frequency $j$
being excited to occupation number $m_j$ in response to a
temperature $T=1/\beta$.

The grand canonical partition function for a single stack of fuzzy
spheres is thus
\begin{equation}
  Z_k(\beta,\nu) = \sum_{N,E=0}^\infty e^{-\nu k N}\, e^{-\beta E}\,
  p_N(E)~,
\end{equation}
where $k$ is the number of units of longitudinal momentum of each
sphere in the stack.  Using Eq.~(\ref{eq-zpn}) this
becomes~\cite{LinMal05}
\begin{equation}
  Z_k(\beta,\nu) = \prod_{n=0}^\infty\frac{1}{1-e^{-\nu k}\,
    e^{-\beta n}}~.
\end{equation}

A general vacuum contains stacks of fuzzy spheres of arbitrary
momenta, described by the product of these partition functions
(analogous to the product of partition functions of harmonic
oscillators of different frequencies):
\begin{equation}
Z(\beta,\nu) = \prod_{k=1}^\infty Z_k(\beta,\nu)~.
\label{eq-zbm}
\end{equation}
For small $\beta$ and $\nu$ this evaluates to
\begin{equation}
\log Z(\beta,\nu) \approx \frac{c}{\beta\nu}~,
\end{equation}
with $c\approx 1.20206$.  

The entropy is obtained from the integral transform
\begin{equation}
e^{S_{\rm p}(N,E)} = \int d\nu\, d\beta\, e^{\nu N+\beta E}\, Z(\beta,\nu)~,
\end{equation}
where in the saddlepoint approximation
\begin{equation}
N=\frac{c}{\beta\nu^2}~,~~~ E=\frac{c}{\beta^2 \nu}~.
\label{eq-sadp}
\end{equation}
This yields a density of states $\frac{1}{2} (\frac{c}{N^2E^2})^{1/3}
\exp[3(cNE)^{1/3}]$, so that to leading order~\cite{LinMal05}
\begin{equation}
S_{\rm p}(N,E) = 3(cNE)^{1/3}~.
\label{eq-fent}
\end{equation}

Using Eq.~(\ref{eq-sadp}), the assumption of small $\beta$ and $\nu$
translates into the regime of validity
\begin{equation}
\sqrt{N}\ll E\ll N^2~.
\end{equation}
A lower crossover is at $E^2\sim N$, when the entropy becomes
completely dominated by the number of vacua (the partitions of $N$).
We are interested in the higher crossover at $E\sim N^2$.  At this
point the entropy becomes dominated by the states in the $X=0$ sector.
From Eq.~(\ref{eq-fent}) we find again that the entropy at this
crossover energy is of order $N$:
\begin{equation}
S_{\rm p, cross} \sim N~.
\label{eq-allcr}
\end{equation}

To approach the crossover from the other side ($E\gg N^2$), we
evaluate the partition function (\ref{eq-zbm}) for $\nu\gg 1$:
\begin{equation}
\log Z(\beta,\nu) \approx \frac{1}{\beta e^\nu}~,
\end{equation}
with saddlepoint
\begin{equation}
E=\frac{1}{\beta^2 e^\nu}~,~~~N = \frac{1}{\beta e^\nu}~.
\end{equation}
Hence the asymptotic behavior of the entropy at very high energies is
\begin{equation}
S_{\rm p} = 2N+N\log\frac{E}{N^2}~.
\end{equation}
For $E\sim N^2$, this matches up with Eq.~(\ref{eq-fent}) at the
crossover entropy (\ref{eq-allcr}).

\section{Assumptions and implications}
\label{sec-discuss}

Our analysis shows that the Bekenstein bound, in the DLCQ form
$S\lesssim N$, is satisfied and approximately saturated in the plane
wave matrix model under the following assumptions:
\begin{enumerate}
\item{The protected states correctly estimate the entropy going into
    the Bekenstein bound, at least for energies approaching the
    crossover scale $E=N^2$ from below.}
\item{For energies above the crossover the matrix model does not
    describe a weakly gravitating system.}
\end{enumerate}
In this section we discuss why these assumptions are needed, point out
that the first lends plausibility to the second, and speculate about
their implications for the matrix model.

Applicability of the Bekenstein bound requires a good semiclassical
background.  This is why we have studied the spectrum for small values
of $\mu$, where curvatures are small.  Because the matrix model is
strongly coupled there, we were only able to discuss protected states.

But a weak background is not enough.  The states included in the
Bekenstein bound must themselves be weakly gravitating, in the sense
that they are incapable of focussing light significantly over a
distance set by their own spatial size~\cite{Bou03}.

Some protected states, even in a weakly curved background, may well
have large backreaction.  (In the context of AdS, an example are
single gravitons with energy above the Planck energy.)  Thus, not all
protected states necessarily contribute to the entropy in the
Bekenstein bound.  On the other hand, there may be some unprotected
states (in the AdS analogy, say, multiple weak gravitons) which have
small backreaction and do contribute.

Such effects will not be important if they modify the entropy only by
factors of order one, or only far from the crossover energy.  But if
they are significant near $E\sim N^2$, they can affect one or both of
our conclusions (that the Bekenstein bound is satisfied, and that it
is saturated).  This is why the first assumption is needed.  Without
the second assumption, the Bekenstein bound would be violated by
states with $E\gg N^2$.

We are hopeful that the validity of both assumptions can be clarified
thanks to recent progress in the physical understanding of the plane
wave states and their geometry at strong coupling.  Excitations of M5
branes appear in the spectrum of protected states~\cite{M5}, and their
multiparticle states are likely to play a role in this analysis.
Since the geometry near a single M5 brane is not smooth, it will be
interesting to consider solutions with coincident M5 brane
configurations~\cite{LinMal05} in this context.  We leave this to
future work.

If the first assumption is correct, it will have an interesting
implication that makes the second assumption more plausible: The
crossover behavior at $E\sim N^2$ (in units of $\mu$), which is quite
well understood at weak coupling, will persist at strong coupling.
This may seem surprising: When $\mu\ll M_{\rm Pl}$, why should the
characteristic scale continue to be $\mu$, rather than, say, the
Planck scale?  Consider a scattering problem with impact parameter
much shorter than the transverse curvature radius, $(R/\mu)^{1/2}$
(which is large in Planck units in the strong coupling regime).  Such
processes should be described as in the flat space matrix model.
Therefore, $\mu$ should be dynamically irrelevant, and all interesting
scales should arise from the Planck mass and powers of $N$.

However, this logic applies only to short timescales.  In the flat
space case ($\mu=0$ exactly)  two gravitons that scatter can move off
to infinity uninhibited by the quartic interaction  since off-diagonal
excitations are frozen out at large separation.  But gravitons that
scatter in the plane wave will eventually come to notice that they
live in a confining background and are really in a bound state.  This
takes a time of order $1/\mu$, which diverges in the $\mu\to 0$ limit.
None of these bound states survive at $\mu=0$, so we do not expect the
BFSS limit to be smooth in this sense.  This argument suggests that
$\mu$ remains an important scale in the microcanonical ensemble at all
nonzero values of $\mu$  no matter how small.

Phase transitions in Yang-Mills theory have been analyzed in detail in
Ref.~\cite{AhaMar03}.  In terms of their classification, our analysis
suggests that the plane wave matrix model will {\em not\/} behave
like, say, ordinary $d=4$ Yang-Mills theory, whose transition
temperature can be set by two different scales. ($T_{\rm crit}\sim
1/R$ for compactification on a small three-manifold of size $R^3$, but
$T\sim\Lambda_{\rm QCD}$ for $R\gg 1/\Lambda_{\rm QCD}$, when coupling
is strong.)  Rather the matrix model should behave like $d=4$, ${\cal
  N}=4$ super-Yang-Mills theory on a three-sphere, whose phase
structure is expected to be the same at strong and weak coupling. (The
critical temperature is of order $1/R$ and the crossover energy is
$N^2/R$, where $R$ is the size of the $S^3$.)

It is interesting to ask about the gravity dual to the matrix model
above the crossover energy.  The deconfinement phase transition of
${\cal N}=4$, $d=4$ super-Yang-Mills theory is related via AdS/CFT to
the Hawking-Page transition in AdS~\cite{Wit98a,Wit98b}.  More
generally, it has been suggested that the high energy phase of a
large-$N$ Yang-Mills theory always corresponds to the presence of a
black object in the dual closed string background~\cite{AhaMar03}.
This amounts to a nonperturbative modification of the background.

Here we are dealing with an M-theory background, but it is tempting to
speculate that it becomes similarly modified.  Conceivably, the bulk
dual of the crossover could be more drastic: the gravity
interpretation may break down entirely.  Unlike for AdS, no black hole
or black string solutions asymptotic to the eleven-dimensional plane
wave are known.  Moreover, it is doubtful that the canonical ensemble
can be defined any more sensibly than it can for flat
space.\footnote{We thank M.~Van Raamsdonk for stressing this point to
  us.}  (To put flat space at a finite temperature requires nonzero
constant energy density and thus infinite energy; in a theory with
gravity this invalidates the background.)

In AdS/CFT gravity turns off for $N\to\infty$, where the crossover
becomes a sharp phase transition.  In this limit there is no longer a
good gravity description above the critical temperature.  Unlike AdS,
the M-theory limit on the plane wave {\em requires\/} $N\to\infty$.
This is another reason to suspect that there may be no sensible
eleven-dimensional gravity dual beyond the crossover energy.

For the purposes of obtaining a cutoff on the entropy entering the
Bekenstein bound, it only matters that such states can no longer be
described as small perturbations of the plane wave background.  Any of
the scenarios we have described---black hole formation, or a complete
breakdown of the geometric interpretation---would guarantee this.

\acknowledgments

We would like to thank M.~Aganagic, O.~DeWolfe, B.~Freivogel,
C.~Keeler, H.~Lin, J.~Maldacena, S.~Shenker, B.~Tweedie and especially
M.~Van Raamsdonk for helpful discussions.  This work was supported by
the Berkeley Center for Theoretical Physics, by a CAREER grant of the
National Science Foundation, and by DOE grant DE-AC03-76SF00098.

\appendix

\section{The $N=1$ sector}
\label{sec-n1}

In the geometric regime the matrix model is strongly coupled, except
in the case $N=1$ when all interaction terms vanish.  This sector is
present for all values of $N$ since $U(N)=U(1)\times SU(N)$.  It
describes the decoupled dynamics of the center of mass.  In this
appendix we discuss the spectrum of the $U(1)$ sector.  We will argue
that it does not contribute to the entropy in the Bekenstein bound
because all of its states are gauge copies of each other under
diffeomorphisms.

The Hamiltonian and creation operators for the $N=1$ sector can be obtained
from Eqs.~(\ref{eq-ham}) and (\ref{eq-ops}) as a special case.  It is
the Hamiltonian for a particle in a nine-dimensional harmonic
oscillator, namely a graviton of longitudinal momentum $1/R$ in
linearized 11D supergravity~\cite{KimYos03}.  Its frequency is set by
the curvature parameter of the plane wave, $\mu$, and its effective
mass is given by the longitudinal momentum, $1/R$.  Hence, the
characteristic length scale of the oscillator is $(R/\mu)^{1/2}$.  The
spectrum is given by infinite towers of bosonic states generated by
the creation operators $A^\dagger$, with an additional multiplicity of
$2^8$ from the fermionic operators $\psi^\dagger$.

Application of the Bekenstein bound requires not only that the
background be weakly curved, but also that the backreaction of the
system we study be negligible.  In particular, the light-sheet in the
$x^-$ direction should not focus much over a distance $2\pi R$.  The
total area loss of the light-sheet is $N$~\cite{Bou03}.  This
corresponds to negligible focussing only if the transverse area of the
system (here, the graviton) is much larger than $N$:
\begin{equation}
A\gg N~.
\label{eq-an}
\end{equation}
In the ground state, the graviton occupies a transverse area of order
$(R/\mu)^{9/2}$, so weak backreaction requires
\begin{equation}
\mu/R\ll 1~.
\label{eq-br1}
\end{equation}
This condition is automatically satisfied in a weakly curved
background; see Sec.~\ref{sec-pwmm}.

The spread of the graviton wave function in the transverse directions
will be of order $\sqrt{RE/\mu}$.  The area loss along $x^-$ is $N$
independently of $E$, so the backreaction becomes weaker for excited
states.  The point is that the spreading of the wave function over a
nine-dimensional area overcompensates for the larger energy leading
to a decrease in energy density and in backreaction.  Hence, the
spectrum will not be truncated due to large backreaction.

Under the criteria offered so far, it would appear that all states of
the nine-dimensional oscillator contribute to the entropy as long as
we choose $\mu/R\ll 1$.  Their number is infinite\footnote{The entropy
  grows with energy as $S_{N=1}(E) \sim 9 \log E$, but there is no
  cutoff on $E$.}, so the Bekenstein bound would seem to be violated
already for $N=1$.

However, this is clearly wrong for the simple reason that the states
we have considered are all the same.  It is easiest to see this in the
(overcomplete) basis of coherent states.  All coherent states can be
mapped to the ground state by an $SU(4|2)$ transformation, i.e., by a
symmetry transformation that leaves the form (\ref{eq-metric}) of the
plane wave metric invariant.  The required generators~\cite{BlaOlo02}
commute with $\partial_-$, so the (discrete) longitudinal momentum is
not affected by this transformation.  Hence, the Bekenstein bound is
trivially satisfied for $N=1$.

The same argument will apply to the center-of-mass motion for larger
values of $N$.  In the matrix model this corresponds to the $U(1)$
factor that decouples from the $SU(N)$ degrees of freedom.  We discard
the $U(1)$ states for the same reason that we would not count
different boosts of the same system as distinct bound states in flat
space.

\section{Protected representations}
\label{sec-protected}

Here we give a brief summary of the form and properties of
supersymmetrically protected states in the plane wave matrix model.
For more details see, e.g., Refs.~\cite{DSV2,M5,LinMal05}.

The symmetry algebra of the eleven-dimensional plane wave is a basic
classical Lie superalgebra
\begin{equation}
SU(4|2) \supset SU(4) \times SU(2) \times U(1)_H \sim SO(6) \times 
SO(3) \times U(1)_H~.
\end{equation}
The bosonic subalgebra $SO(6) \times SO(3)$ describes the symmetry of
the nine-dimensional transverse plane, and the Hamiltonian is the
$U(1)_H$ hypercharge.  One can also think of this symmetry group as
arising in the Penrose limit of AdS/CFT.  On the AdS side, the Penrose
limit of $AdS_7 \times S^4$ gives the plane wave.  On the CFT side,
the corresponding contraction of the $\mathcal{N}=4$ superconformal
group $SU(2,2|4)$ gives the supergroup $SU(4|2)$.

At $\mu\to\infty$, all superrepresentations in the plane wave matrix
model are tensor representations.  Hence they are described by Young
tableaux.  To distinguish a supertableau from a bosonic tableau,
slashed boxes are used.  Any $SU(4|2)$ superrepresentation can be
completely decomposed into representations of the bosonic subgroup
$SU(4) \times SU(2)$.  For example, the superrepresentation that is
the sole building block of the entire $U(1)$ spectrum has the
following decomposition
\begin{equation} 
\label{eq-superdecomp}
{\tiny \yngSLASH(1,1)} \; = \;
\left( \tableau{1 1}, 1 \right) \; \oplus \; \;
\left( \tableau{1}, \tableau{1} \right) \; \; \oplus \;
\left( 1, \tableau{2} \right)~.
\end{equation}

Conversely, starting with a highest weight bosonic representations
$\vert \psi \rangle$, the full superrepresentation can be recovered by
acting with the $2^8$ combinations of fermionic lowering operators.
If the resulting states are all independent, the superrepresentation
is called ``typical''; otherwise, ``atypical''.  

A special set of multiplets are nonperturbatively protected from
receiving corrections to their energy as the parameter $\mu$ of the
matrix model is modified.  The Young supertableaux of these ``doubly
atypical'' representations have two rows of equal length
\begin{equation}
\label{eq-DAform}
\cdot \, , \; {\tiny \yngSLASH(1,1) } \, , \; 
{\tiny \yngSLASH(2,2) } \, , \; \ldots, \; 
\underbrace{ {\tiny \yngSLASH(6,6) } }_n \, , \; \ldots
\end{equation}
and dimension
\begin{equation}
\rm{dim}_{\rm{DA}}(n)=\frac{1}{3}\left( 4n^4+16n^3+20n^2+8n+3 \right)~.
\end{equation}
Their bosonic decomposition is (for $n \geq 2$)
\begin{eqnarray}
\label{eq-DAdecomp}
\!\!\!\!\!\!\!\!\!\!\!\!\!\!\!
 \underbrace{ \tiny{ \yngSLASH(6,6) } }_n \; &=& \;
\Big( \underbrace{ \tableau{6 6}}_n \, , 
     1 \Big)_{{\bf \frac{1}{4}}\,\rm{BPS} } \; \oplus \;
\Big( \underbrace{ \tableau{6 5}}_n \, , 
     \tableau{1} \Big)_{{\bf \frac{1}{8}}\,\rm{BPS}} \; \oplus \;
\Big( \underbrace{ \tableau{5 5}}_{n-1} \, , 
     \tableau{2} \Big)_{{\bf \frac{1}{8}}\,\rm{BPS}} \nonumber \\
&\oplus& \; \Big( \underbrace{ \tableau{6 4}}_{n} \, , 
     \tableau{1 1} \Big)_{{\bf \frac{1}{4}}\,\rm{BPS}} \; \oplus \;
\Big( \underbrace{ \tableau{5 4}}_{n-1} \, , 
     \tableau{2 1} \Big)_{{\bf \frac{1}{8}}\,\rm{BPS}} \; \oplus \;
\Big( \underbrace{ \tableau{4 4}}_{n-2} \, , 
     \tableau{2 2} \Big)_{{\bf \frac{1}{4}}\,\rm{BPS}}~,
\end{eqnarray}
where, following Ref.~\cite{DSV2}, each bosonic subrepresentations has
been labeled by the fraction of the 32 supersymmetries it preserves.
The energy of all constituent states is of order $E \sim n$.

The nonstandard supersymmetry algebra on the plane wave allows for BPS
states with nonzero energy.  This follows from the commutation
relations between the supersymmetry generators and the Hamiltonian
$H$. Schematically,
\begin{eqnarray}
\left\{ Q^{\dagger}, Q \right\} &\sim& H - \mu M^{ij} - \mu M^{ab}~,
\nonumber \\
\left[ H, Q \right] &\sim& \mu Q~,
\end{eqnarray}
where the $SO(3)$, $SO(6)$ rotation generators $M^{ij}$, $M^{ab}$
allow for positive definite $H$ when $\left\{ Q^{\dagger}, Q
\right\}=0$.

\bibliographystyle{board}
\bibliography{all}

\begin{thebibliography}{10}

\bibitem{Tho93}
G.~'t~Hooft: {\em Dimensional reduction in quantum gravity\/}, gr-qc/9310026.

\bibitem{Sus95}
L.~Susskind: {\em The world as a hologram\/}. J. Math. Phys. {\bf 36}, 6377
  (1995), hep-th/9409089.

\bibitem{CEB2}
R.~Bousso: {\em Holography in general space-times\/}. JHEP {\bf 06}, 028
  (1999), hep-th/9906022.

\bibitem{CEB1}
R.~Bousso: {\em A covariant entropy conjecture\/}. JHEP {\bf 07}, 004 (1999),
  hep-th/9905177.

\bibitem{Mal97}
J.~Maldacena: {\em The large {$N$} limit of superconformal field theories and
  supergravity\/}. Adv. Theor. Math. Phys. {\bf 2}, 231 (1998), hep-th/9711200.

\bibitem{SusWit98}
L.~Susskind and E.~Witten: {\em The holographic bound in {A}nti-de~{S}itter
  space\/}, {h}ep-th/9805114.

\bibitem{PeePol98}
A.~W. Peet and J.~Polchinski: {\em {UV}/{IR} relations in {AdS} dynamics\/}.
  Phys. Rev. D {\bf 59}, 065011 (1999), hep-th/9809022.

\bibitem{BFSS}
T.~Banks, W.~Fischler, S.~H. Shenker and L.~Susskind: {\em {M} theory as a
  matrix model: {A} conjecture\/}. Phys. Rev. D {\bf 55}, 5112 (1997),
  hep-th/9610043.

\bibitem{BMN}
D.~Berenstein, J.~M. Maldacena and H.~Nastase: {\em Strings in flat space and
  pp-waves from {N} = 4 super {Y}ang {M}ills\/}. JHEP {\bf 04}, 013 (2002),
  hep-th/0202021.

\bibitem{Bek81}
J.~D. Bekenstein: {\em A universal upper bound on the entropy to energy ratio
  for bounded systems\/}. Phys. Rev. D {\bf 23}, 287 (1981).

\bibitem{FMW}
E.~E. Flanagan, D.~Marolf and R.~M. Wald: {\em Proof of classical versions of
  the {B}ousso entropy bound and of the {G}eneralized {S}econd {L}aw\/}. Phys.
  Rev. D {\bf 62}, 084035 (2000), hep-th/9908070.

\bibitem{Bou03}
R.~Bousso: {\em Light-sheets and {B}ekenstein's bound\/}. Phys. Rev. Lett. {\bf
  90}, 121302 (2003), hep-th/0210295.

\bibitem{Bou04}
R.~Bousso: {\em Flat space physics from holography\/}. JHEP {\bf 05}, 050
  (2004), hep-th/0402058.

\bibitem{Pag00c}
D.~N. Page: {\em Defining entropy bounds\/}  (2000), hep-th/0007238.

\bibitem{Bek00b}
J.~D. Bekenstein: {\em On {P}age's examples challenging the entropy bound\/}
  (2000), gr-qc/0006003.

\bibitem{Bou03a}
R.~Bousso: {\em Bound states and the {B}ekenstein bound\/}. JHEP {\bf 02}, 025
  (2004), hep-th/0310148.

\bibitem{MarRoi04}
D.~Marolf and R.~Roiban: {\em Note on bound states and the {B}ekenstein
  bound\/}. JHEP {\bf 08}, 033 (2004), hep-th/0406037.

\bibitem{MarMin03}
D.~Marolf, D.~Minic and S.~F. Ross: {\em Notes on spacetime thermodynamics and
  the observer-dependence of entropy\/}. Phys. Rev. {\bf D69}, 064006 (2004),
  hep-th/0310022.

\bibitem{Pag04}
D.~N. Page: {\em Hawking radiation and black hole thermodynamics\/}. New J.
  Phys. {\bf 7}, 203 (2005), hep-th/0409024.

\bibitem{Bek04}
J.~D. Bekenstein: {\em How does the entropy/information bound work ?\/}
  (2004), quant-ph/0404042.

\bibitem{Bou03b}
R.~Bousso: {\em Harmonic resolution as a holographic quantum number\/}. JHEP
  {\bf 03}, 054 (2004), hep-th/0310223.

\bibitem{BouFla03}
R.~Bousso, E.~E. Flanagan and D.~Marolf: {\em Simple sufficient conditions for
  the generalized covariant entropy bound\/}. Phys. Rev. D {\bf 68}, 064001
  (2003), hep-th/0305149.

\bibitem{StrTho03}
A.~Strominger and D.~M. Thompson: {\em A quantum {B}ousso bound\/}. Phys. Rev.
  {\bf D70}, 044007 (2004), hep-th/0303067.

\bibitem{Sus97}
L.~Susskind: {\em Matrix theory black holes and the {G}ross-{W}itten
  transition\/}  (1997), hep-th/9805115.

\bibitem{DSV1}
K.~Dasgupta, M.~M. Sheikh-Jabbari and M.~Van~Raamsdonk: {\em Matrix
  perturbation theory for {M}-theory on a pp-wave\/}. JHEP {\bf 05}, 056
  (2002), hep-th/0205185.

\bibitem{AhaMar03}
O.~Aharony, J.~Marsano, S.~Minwalla, K.~Papadodimas and M.~Van~Raamsdonk: {\em
  The {H}agedorn / deconfinement phase transition in weakly coupled large {N}
  gauge theories\/}. Adv. Theor. Math. Phys. {\bf 8}, 603 (2004),
  hep-th/0310285.

\bibitem{HawPag83}
S.~W. Hawking and D.~N. Page: {\em Thermodynamics of black holes in {A}nti-de
  {S}itter space\/}. Commun. Math. Phys. {\bf 87}, 577 (1983).

\bibitem{Wit98a}
E.~Witten: {\em {A}nti-de~{S}itter space and holography\/}. Adv. Theor. Math.
  Phys. {\bf 2}, 253 (1998), hep-th/9802150.

\bibitem{Sun00}
B.~Sundborg: {\em The {H}agedorn transition, deconfinement and {N} = 4 sym
  theory\/}. Nucl. Phys. {\bf B573}, 349 (2000), hep-th/9908001.

\bibitem{AhaMar04}
O.~Aharony, J.~Marsano, S.~Minwalla and T.~Wiseman: {\em Black hole - black
  string phase transitions in thermal 1+1 dimensional supersymmetric
  {Y}ang-{M}ills theory on a circle\/}. Class. Quant. Grav. {\bf 21}, 5169
  (2004), hep-th/0406210.

\bibitem{AhaMar05a}
O.~Aharony, J.~Marsano, S.~Minwalla, K.~Papadodimas and M.~Van~Raamsdonk: {\em
  A first order deconfinement transition in large {N} {Y}ang- {M}ills theory on
  a small {S}**3\/}. Phys. Rev. {\bf D71}, 125018 (2005), hep-th/0502149.

\bibitem{AhaMar05b}
O.~Aharony, J.~Marsano, S.~Minwalla, K.~Papadodimas, M.~Van~Raamsdonk and
  T.~Wiseman: {\em The phase structure of low dimensional large {N} gauge
  theories on tori\/}  (2005), hep-th/0508077.

\bibitem{LiMar98}
M.~Li, E.~J. Martinec and V.~Sahakian: {\em Black holes and the {SYM} phase
  diagram\/}. Phys. Rev. {\bf D59}, 044035 (1999), hep-th/9809061.

\bibitem{HadRam04}
S.~Hadizadeh, B.~Ramadanovic, G.~W. Semenoff and D.~Young: {\em Free energy and
  phase transition of the matrix model on a plane-wave\/}. Phys. Rev. {\bf
  D71}, 065016 (2005), hep-th/0409318.

\bibitem{Sem04}
G.~W. Semenoff: {\em Matrix model thermodynamics\/}  (2004), hep-th/0405107.

\bibitem{FurSch03}
K.~Furuuchi, E.~Schreiber and G.~W. Semenoff: {\em Five-brane thermodynamics
  from the matrix model\/}  (2003), hep-th/0310286.

\bibitem{BouMin05a}
R.~Bousso and A.~L. Mints: {\em Decoding the matrix: Coincident membranes on
  the plane wave\/}  (2005), hep-th/0510121.

\bibitem{DSV2}
K.~Dasgupta, M.~M. Sheikh-Jabbari and M.~Van~Raamsdonk: {\em Protected
  multiplets of {M}-theory on a plane wave\/}. JHEP {\bf 09}, 021 (2002),
  hep-th/0207050.

\bibitem{BalDeb05}
V.~Balasubramanian, J.~de~Boer, V.~Jejjala and J.~Simon: {\em The library of
  {B}abel: On the origin of gravitational thermodynamics\/}  (2005),
  hep-th/0508023.

\bibitem{TraMur03}
M.~N. Tran, M.~V.~N. Murthy and R.~K. Bhaduri: {\em On the quantum density of
  states and partitioning an integer\/}. Annals Phys. {\bf 311}, 204 (2004),
  math-ph/0309020.

\bibitem{LinMal05}
H.~Lin and J.~Maldacena: {\em Fivebranes from gauge theory\/}  (2005),
  hep-th/0509235.

\bibitem{M5}
J.~Maldacena, M.~M. Sheikh-Jabbari and M.~Van~Raamsdonk: {\em Transverse
  fivebranes in matrix theory\/}. JHEP {\bf 01}, 038 (2003), hep-th/0211139.

\bibitem{Wit98b}
E.~Witten: {\em {A}nti-de~{S}itter space, thermal phase transition, and
  confinement in gauge theories\/}. Adv. Theor. Math. Phys. {\bf 2}, 505
  (1998), hep-th/9803131.

\bibitem{KimYos03}
T.~Kimura and K.~Yoshida: {\em Spectrum of eleven-dimensional supergravity on a
  pp-wave background\/}. Phys. Rev. {\bf D68}, 125007 (2003), hep-th/0307193.

\bibitem{BlaOlo02}
M.~Blau and M.~O'Loughlin: {\em Homogeneous plane waves\/}. Nucl. Phys. {\bf
  B654}, 135 (2003), hep-th/0212135.

\end{thebibliography}
\end{document}